\documentclass[aps,prd,notitlepage,showpacs,nofootinbib]{revtex4-1}

\usepackage{graphicx}

\usepackage[utf8]{inputenc} 

\usepackage{amsmath}
\usepackage{amssymb}
\usepackage{float}
\usepackage{comment}

\def\gsim{\raise0.3ex\hbox{$\;>$\kern-0.75em\raise-1.1ex\hbox{$\sim\;$}}}
\def\lsim{\raise0.3ex\hbox{$\;<$\kern-0.75em\raise-1.1ex\hbox{$\sim\;$}}}
\def\lfv{lepton flavour violation }

\newcommand{\sm}{{Standard Model }}

\usepackage{soul}

\usepackage[usenames,dvipsnames]{color}
\usepackage[colorinlistoftodos]{todonotes}

\definecolor{mightnightblue}{RGB}{25,25,112}

\definecolor{brown}{rgb}{0.59, 0.29, 0.0}

\def\lfv{lepton flavour violation }

\def\21{$\mathrm{SU(2)_L \otimes U(1)_Y}$}
\def\lfv{lepton flavour violation }

\def\sm{standard model }

\def\Q{\hbox{$\cal Q$ }}

\newcommand{\AddrAHEP}{AHEP Group, Institut de F\'{i}sica Corpuscular --
  C.S.I.C./Universitat de Val\`{e}ncia, Parc Cientific de Paterna.\\
  C/Catedratico Jos\'e Beltr\'an, 2 E-46980 Paterna (Val\`{e}ncia) - SPAIN}

\newcommand{\UdeA}{Instituto de F\'\i sica, Universidad de Antioquia,
  Calle 70 No. 52-21, Apartado Aéreo 1226, Medellín, Colombia.}
  
\usepackage{hyperref}
 \hypersetup{
     colorlinks=true,
     linkcolor= Black,
     citecolor=mightnightblue,
     urlcolor=mightnightblue
     } 

\begin{document}
\title{\color{BrickRed}Bound-state dark matter and Dirac neutrino mass}
\author{M. Reig~${}^1$}\email{mario.reig@ific.uv.es}
\author{D. Restrepo~${}^2$}\email{restrepo@udea.edu.co}
\author{J. W. F. Valle~${}^1$} \email{valle@ific.uv.es, URL: de
  http://astroparticles.es/} 
\author{O. Zapata~${}^2$}\email{oalberto.zapata@udea.edu.co}

\affiliation{$^1$~\AddrAHEP}
\affiliation{$^2$~\UdeA}


\begin{abstract}

  We propose a simple theory for the idea that cosmological dark
  matter (DM) may be present today mainly in the form of stable
  neutral hadronic thermal relics.
  In our model neutrino masses arise radiatively from the exchange of
  colored DM constituents, giving a common origin for both dark matter
  and neutrino mass.
  The exact conservation of $B-L$ symmetry ensures dark matter
  stability and the Dirac nature of neutrinos.
  The theory can be falsified by dark matter nuclear recoil direct
  detection experiments, leading also to possible signals at a next
  generation hadron collider.

  \end{abstract}


\maketitle

\section{Introduction}

The common lore concerning particle dark matter candidates has long
been that they must be electrically neutral and carry no color.
This view has been challenged by the authors in
Ref.~\cite{DeLuca:2018mzn}, who suggested that dark matter (DM) may be
the lightest hadron made of two stable color octet Dirac fermions \Q
with mass below $10~\text{TeV}$.
The interest of this idea may go well beyond QCD, since analogous
bound-state DM candidates emerge in models with a new confining
hypercolor interaction~\cite{Reig:2017nrz}.

We argue that neutrino mass and cosmological dark matter may have a
common origin, with the underlying DM physics acting as messenger of
neutrino mass
generation~\cite{Ma:2006km,Hirsch:2013ola,Merle:2016scw,Bonilla:2016diq}.
We propose a simple implementation of the bound-state dark matter
scenario in which DM constituents induce calculable neutrino mass at
the radiative level.
Dark matter is a QCD bound state $\mathcal{Q}\mathcal{Q}$ stabilized
by the same conserved $B-L$ symmetry associated to the Dirac nature of
neutrinos~\cite{Chulia:2016ngi}.

In addition to the heavy Dirac fermion, our model introduces extra
scalars, in order to ensure that at least two neutrino masses are
nonzero, required by the neutrino oscillation data in order to account
both for solar and atmospheric mass scales~\cite{deSalas:2017kay}.
This simple picture can account for current neutrino oscillation and
dark matter phenomena, and can be falsified relatively soon, in
nuclear recoil studies at XENON1T~\cite{Aprile:2015uzo}.
Moreover, the extra colored states, including scalar bosons, may
lead to new phenomena at a next generation hadron collider.

\section{The model}
\label{sec:model}

As a theory preliminary, we recall that, within the type-I seesaw
mechanism with a single right-handed neutrino, two neutrinos remain
massless after the seesaw~\cite{Schechter:1980gr}.
This degeneracy is lifted by calculable loop corrections~\footnote{An
  analogous situation happens in supersymmetry with bilinear breaking
  of R parity~\cite{Hirsch:2000ef,diaz:2003as,hirsch:2004he}, which
  induces the solar scale at one-loop, once atmospheric scale is taken
  as tree level input.}.
Here we propose a variant radiative seesaw scheme, in which a single
colored fermion Dirac messenger \Q is introduced, charged under the
$\operatorname{U}(1)_{B-L}$ symmetry, plus two sets of colored
scalars, labeled by $a=1,2$, see table~\ref{tab:bmlnur}.
\begin{table}[!h]
  \centering
  \begin{tabular}{|l|c|l|c|}
\hline
    Particles& $\operatorname{U}(1)_{B-L}$ & $\left(\operatorname{SU}(3)_c ,\operatorname{SU}(2)_L \right)_Y$& $Z_2$\\\hline
$Q_{i}=\begin{pmatrix}u_L& d_L\end{pmatrix}_i^{\operatorname{T}}$ & $+1/3$  & $ \left(\mathbf{3},\mathbf{2}  \right)_{1/6}$ & $+$ \\
$\overline{u_{Ri}}$ & $-1/3$  & $ \left(\overline{\mathbf{3}},\mathbf{1}  \right)_{-2/3}$ & $+$ \\
$\overline{d_{Ri}}$ & $-1/3$  & $ \left(\overline{\mathbf{3}},\mathbf{1}  \right)_{1/3}$ & $+$ \\ \hline
 $L_i=\begin{pmatrix}\nu_L& e_L\end{pmatrix}_i^{\operatorname{T}}$                           & $-1$ & $ \left(\mathbf{1},\mathbf{2}\right)_{-1/2}$ &$+$ \\
$\overline{e_{Ri}}$ & $+1$  & $ \left(\mathbf{1},\mathbf{1}  \right)_{1}$ & $+$ \\ 
 $\overline{\nu_{Ri}}$ & $+1$ & $ \left(\mathbf{1},\mathbf{1}\right)_0$&$-$ \\ \hline
 $\mathcal{Q}_L$    & $-r$ & $ \left(\mathbf{N}_c,\mathbf{1}\right)_0$ &$+$\\
 $ \overline{\mathcal{Q}_R} $ & $r$ & $ \left(\mathbf{N}_c,\mathbf{1}\right)_0$ &$+$\\\hline
 $H$                           &          0  & $ \left(\mathbf{1},\mathbf{2}\right)_{1/2}$ &$+$\\ 
 $\sigma_a$                           &  $1-r$   & $ \left(\mathbf{N}_{c},\mathbf{1}\right)_0$ &$-$\\
 $\eta_a$                        &  $1-r$   & $ \left(\mathbf{N}_c,\mathbf{2}\right)_{1/2}$ &$+$\\
\hline
  \end{tabular}
  \caption{Left-handed fermions and scalars. When the vector-like
    quark \Q is taken as a color octet, $\mathbf{N}_c=\mathbf{8}$, one
    can realize bound-state dark matter scenario proposed
    in~Ref.~\cite{DeLuca:2018mzn}.  }
  \label{tab:bmlnur}
\end{table}

Apart from the right-handed neutrinos, all of the new particles are
colored. For definiteness, we assign them to the octet
$\operatorname{SU}(3)_c$ representation~\footnote{ Notice, however,
  that our neutrino mass discussion also holds if they had different
  $\operatorname{SU}(3)_c$ transformation properties.}.
We also impose that the $B-L$ symmetry holds, together with a $Z_2$
symmetry. The former ensures that \Q has only a Dirac-type mass term,
while the latter forbids the tree-level Dirac neutrino mass terms from
$\overline{ \nu_{Ri}} \widetilde{ H}^\dagger L_j$.
The Lagragian contains the following new terms (summation is
implied over repeated indices, and trace over
$\mathbf{N}_c=\mathbf{8}$ is implicit.)
\begin{align}
  \label{eq:lag}
  \mathcal{L}\supset&-\left[ h^{a}_i  \overline{ \mathcal{Q}_R }\widetilde{\eta}^\dagger_a L_i
  + M_{\mathcal{Q}}\, \overline{\mathcal{Q}_R} \mathcal{Q}_L
  +y^a_i    \overline{\nu_{Ri} }\  \sigma_a^{*} \mathcal{Q}_L  +\text{h.c}\right] \nonumber\\
&-\mathcal{V}(H,\eta_a,\sigma_a ) \,.
\end{align}
The condition $r\ne 1$ forbids Higgs-like Yukawa couplings of
$\eta^\alpha$ to the Standard Model fermions~(for $r=1$ one would need
an additional $Z_2$ symmetry, as in~\cite{Farzan:2012sa}). The new
part of the scalar potential can be cast as
\begin{align}
  \mathcal{V}(H,\eta_a,\sigma_a )&=\mathcal{V}(\eta_a)+\mathcal{V}(\sigma_a)+\mathcal{V}(\eta_a,\sigma_a) \nonumber\\&
 +\mathcal{V}(H,\eta_a)+\mathcal{V}(H,\sigma_a),
\end{align}
where the various terms in the Higgs potential are 
\begin{widetext}
\begin{align}
 \mathcal{V}(\eta_a,\sigma_a)& =\kappa^{ab}\, \operatorname{Tr}\left(\sigma_a \eta_b^{\dagger}\right)H+\lambda_{\sigma \eta}^{abcd} \operatorname{Tr}\left(\sigma_{a}^{*}\sigma_{b} \right) \operatorname{Tr}\left(\eta^{\dagger}_c \eta_d\right) + \tilde{\lambda}_{\sigma \eta}^{abcd} \operatorname{Tr}\left(\sigma_{a}^{*}\eta_{b} \right) \operatorname{Tr}\left(\sigma_c \eta_d^\dagger\right)+\text{h.c.},\\
\mathcal{V}(H,\eta_a)&=\lambda_{3\eta H}^{ab}\left( H^{\dagger}H \right)\operatorname{Tr}\left( \eta_{a}^\dagger  \eta_{b} \right)
   +\lambda_{4\eta H}^{ab}\operatorname{Tr}\left[\left( \eta_{a}^\dagger  H\right)\left(  H^\dagger\eta_{b} \right) \right] + \text{h.c.},\\
\mathcal{V}(H,\sigma_a)&=\lambda_{\sigma H}^{ab} \operatorname{Tr}\left(\sigma_{a}^{*}\sigma_{b}\right) H^{\dagger} H  + \text{h.c.}.
\end{align}
\end{widetext}
Since CP conservation is assumed, the CP-even and CP-odd scalars do
not mix. Moreover, terms like $\left( \eta^\dagger H\right)^2$ are
also forbidden and, as a consequence, the real and imaginary parts of
the scalars with nonzero $B-L$ charges are degenerate.  
Note, however, that the cubic scalar coupling terms $\kappa^{ab}$
breaking the $Z_2$ symmetry softly allow for the mixing between the
$\sigma_a$ and $\eta_a$.  At the end the $4\times4$ mass matrices for
the CP-odd and CP-even scalars are equal, since the $\kappa^{ab}$
terms do not break such a degeneracy.

In order to illustrate the neutrino mass generation mechanism we
consider the following block-diagonal mass matrix for the the CP-even
scalars (in the basis
$S_R=(\eta_{1R}^0,\sigma_{1R}^0,\eta_{2R}^0,\sigma_{2R}^0)^T$): 
\begin{align}
  \mathcal{M}^2_{R} = \begin{pmatrix}
(\mu_{\eta}^{11})^2  & \frac{\kappa_{11}v}{\sqrt{2}}  & 0 & 0 \cr
\frac{\kappa_{11}v}{\sqrt{2}}  &  (\mu_{\sigma}^{11})^2    & 0 & 0 \cr
0 & 0  &  (\mu_{\eta}^{22})^2  &   \frac{\kappa_{22}v}{\sqrt{2}}\cr
0 & 0 & \kappa_{22}\frac{v}{\sqrt{2}}  &  (\mu_{\sigma}^{22})^2 
  \end{pmatrix}.
\end{align}
Here we have used the parametrization 
$\eta_a=(\eta_a^+,(\eta_{aR}^0+i\eta_{aI}^0)/\sqrt{2})^T$,
$\sigma_a=(\sigma_{aR}^0+i\sigma_{aI}^0)/\sqrt{2}$ and
$H=(0,(h+v)/\sqrt{2})^T$, with $h$ denoting the \sm Higgs boson, and
$v=246$ GeV. 
The parameters $\mu_{\eta(\sigma)}^{ab}$ are the quadratic mass terms
after electroweak symmetry breaking in
$\mathcal{V}\left(\eta_a\right)$ and
$\mathcal{V}\left(\sigma_a\right)$.

\begin{figure}
\centering
\includegraphics[scale=0.5]{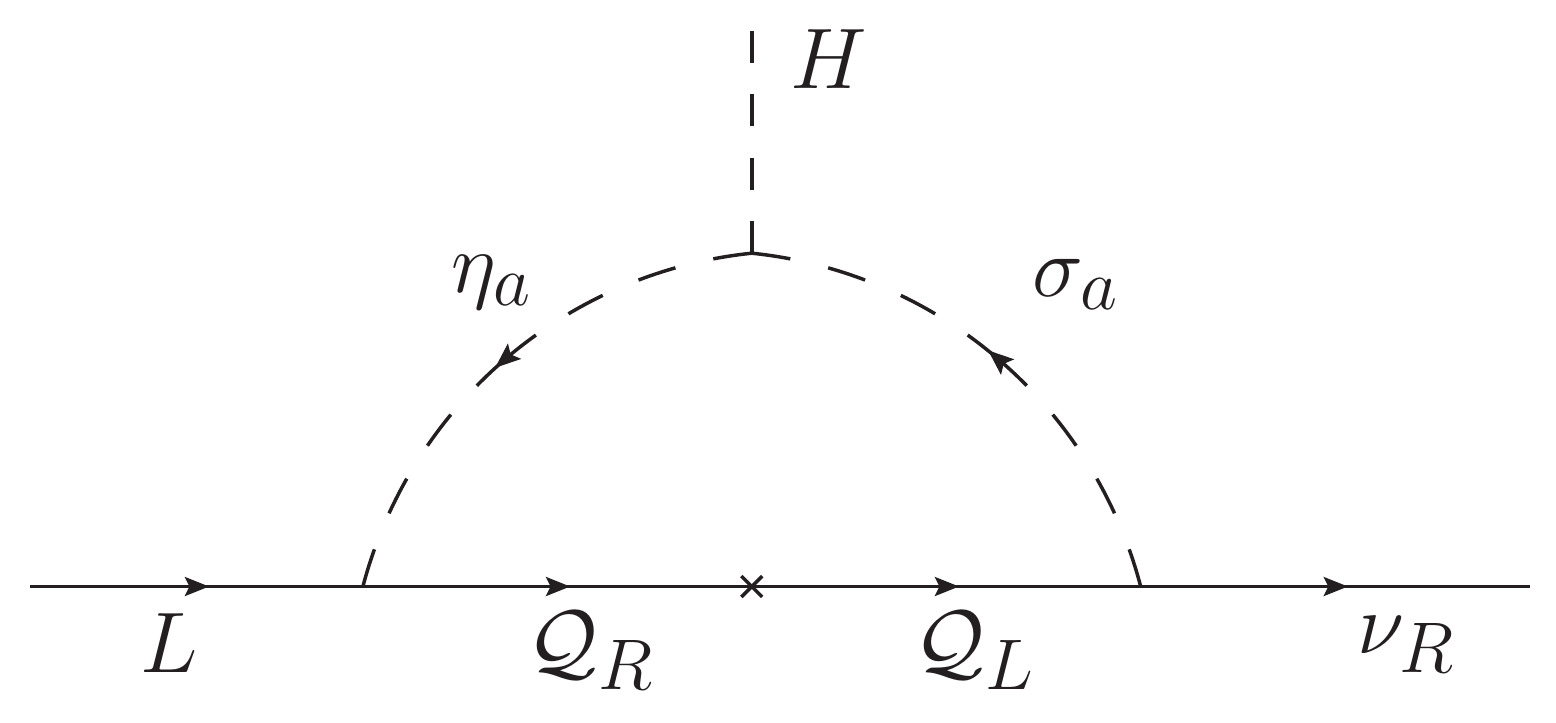}
\caption{$B-L$ flux in the one-loop Dirac neutrino mass.}
\label{fig:oneloop}
\end{figure}

Since the tree-level Dirac mass term is forbidden by symmetry,
calculable neutrino masses are generated at one--loop order, by the
Feynman diagram displayed in~Fig.~\ref{fig:oneloop}.
One finds the following effective mass matrix
\begin{align}
\left(\mathcal{M}_\nu\right)_{ij}=&N_c\frac{M_{\mathcal{Q}}}{64\pi^2}\sum_{a=1}^2h_{i}^a y_{j}^a\frac{\sqrt{2}\kappa_{aa}v}{m^2_{S_{2R}^a}-m^2_{S_{1R}^a}}\nonumber\\
&  \left[F\left(\frac{m^2_{S_{2R}^a}}{M_{\mathcal{Q}}^2}\right)-F\left(\frac{m^2_{S_{1R}^a}}{M_{\mathcal{Q}}^2}\right)\right]+(R\to  I)
\end{align}
where
$F(m_{S_\beta}^2/M_Q^2)=m_{S_\beta}^2\log(m_{S_\beta}^2/M_Q^2)/(m_{S_\beta}^2-M_Q^2)$
and the $\operatorname{SU}(3)_c$ color factor $N_c$ is assumed to be $8$,
since the new particles running in the loop transform as octets.
The four CP-even mass eigenstates are denoted as
$S_{1R}^1,S_{2R}^1,S_{1R}^2,S_{2R}^2$, with a similar notation for the
CP-odd ones.
Notice that the above effective neutrino mass matrix in does not have
the projective rank one structure
$ \left(\mathcal{M}_\nu\right)_{ij}^{\text{tree-level}}\propto h_i
^ay^b_j $
characterizing the Dirac variant~\cite{Chulia:2016ngi} of
the ``incomplete'' tree-level seesaw mechanism, involving the exchange
of a single heavy messenger, called (3,1) in
Ref.~\cite{Schechter:1980gr}~\footnote{Here one would recover the
  rank-one situation characterizing the type-I seesaw in the limit
  where one set of scalars decouples.}.

Here the effective one-loop induced neutrino mass matrix has in
general rank two, as in~\cite{Hehn:2012kz}, implying two non-vanishing Dirac neutrino masses.
As a simple numerical estimate, let's consider the case
$\mu_\eta^{aa}=\mu_\sigma^{aa}\gg\sqrt{2}\kappa^{aa}v$. Taking
$m_{S_{2R}^a}^2-m_{S_{1R}^a}^2=\sqrt{2}\kappa^{aa}v$ and
$m_{S_{2R}^a}^2+m_{S_{1R}^a}^2=2(\mu_\eta^{aa})^2$ one finds
\begin{align}
  \left(\mathcal{M}_\nu\right)_{ij}=&N_c\frac{M_{\mathcal{Q}}}{32\pi^2}\sqrt{2}\kappa^{aa}v\sum_{a=1}^2\frac{h_{i}^a y_{j}^a}{(\mu_\eta^{aa})^2-M_{\mathcal{Q}}^2}\nonumber\\
  &\left[1-\frac{M_{\mathcal{Q}}^2}{(\mu_\eta^{aa})^2-M_{\mathcal{Q}}^2}\log\left(\frac{(\mu_\eta^{aa})^2}{M_{\mathcal{Q}}^2}\right)\right],\nonumber
\end{align}
so that if $(\mu_\eta^{aa})^2\gg M_{\mathcal{Q}}^2$ one has
\begin{align}
 &\left(\mathcal{M}_\nu\right)_{ij}=N_c\frac{M_{\mathcal{Q}}}{32\pi^2}\sqrt{2}v\sum_{a=1}^2\kappa^{aa}\frac{h_{i}^a y_{j}^a}{(\mu_\eta^{aa})^2}\\
 &\sim 0.03\,\mbox{eV}\left(\frac{M_{\mathcal{Q}}}{9.5\, \mbox{TeV}}\right)\left(\frac{\kappa^{aa}}{1\, \mbox{GeV}}\right)\left(\frac{50\,\mbox{TeV}}{\mu_\eta^{aa}}\right)^2\left(\frac{h_{i}^a y_{j}^a}{10^{-6}}\right).\nonumber
\end{align}
One sees that, indeed, small neutrino masses arise naturally by taking
reasonable values for the Yukawa couplings, small value for the soft
breaking parameter $\kappa^{ab}$, as well as sufficiently large values
for the scalar masses.
Notice that the smallness of $\kappa^{ab}$ is natural, as the theory
attains a larger symmetry when $\kappa^{ab}\to 0$, i.e. the smallness of
neutrino mass is symmetry-protected.

In short, concerning neutrino mass generation, our model provides a
colored variant of the one suggested in~\cite{Farzan:2012sa}.
However, although the neutrino mass generation is similar in both
models, the details of the associated physics 
differ substantially.

\section{Bound-state dark matter stability from Dirac neutrinos}
\label{sec:dark-matt-stab}

The Dirac nature of neutrinos may ensure dark matter stability, as
suggested in~\cite{Chulia:2016ngi}.
Here we clone this idea with the proposal that dark matter may be
present today mainly in the form of stable neutral hadronic thermal
relics.
For definiteness we assume DM is a neutral bound-state of colored
constituents, such as $\mathcal{Q}\mathcal{Q}$, where \Q is a
vector-like color octet isosinglet fermion.
It was claimed that a necessary and suficient condition for dark
matter stability in this case is the presence of a global $U(1)_D$
dark baryon number, under which the \Q is
charged~\cite{DeLuca:2018mzn}. 
In our present model construction the role of such apparently
\textit{ad-hoc} symmetry is played by the usual ${B-L}$ symmetry
present in the Standard Model.
In fact, in our model dark matter stability, and the Dirac nature of
the exotic fermion \Q and of the neutrinos are all equivalent, and
result from $B-L$ conservation. 

An adequate thermal relic density of bound-state dark matter requires
the lightest constituent vector-like color octet Dirac fermion,
$\mathcal{Q}$, to have a mass $\approx 9.5\ \text{TeV}$, so that the
$\mathcal{Q}\mathcal{Q}$ hadron weighs approximately
$19\ \text{TeV}$~\cite{DeLuca:2018mzn}.

The set of scalars $\eta^a$, $\sigma^a$ in Fig.~\ref{fig:oneloop} can
be either neutral or charged under ${B-L}$, depending on the baryon
number assignment.
If ${B-L}$ neutral, both $\eta^a$ and $\sigma^a$ are expected to be
unstable, decaying to quarks~\cite{Manohar:2006ga} and two gluons
respectively~\cite{Bityukov:1997dh,Bai:2018jsr}.
In contrast, in the second case, if the lightest octet particle were a
scalar, then it would be strictly stable. 

\begin{figure}[h]
\centering
\includegraphics[scale=0.4]{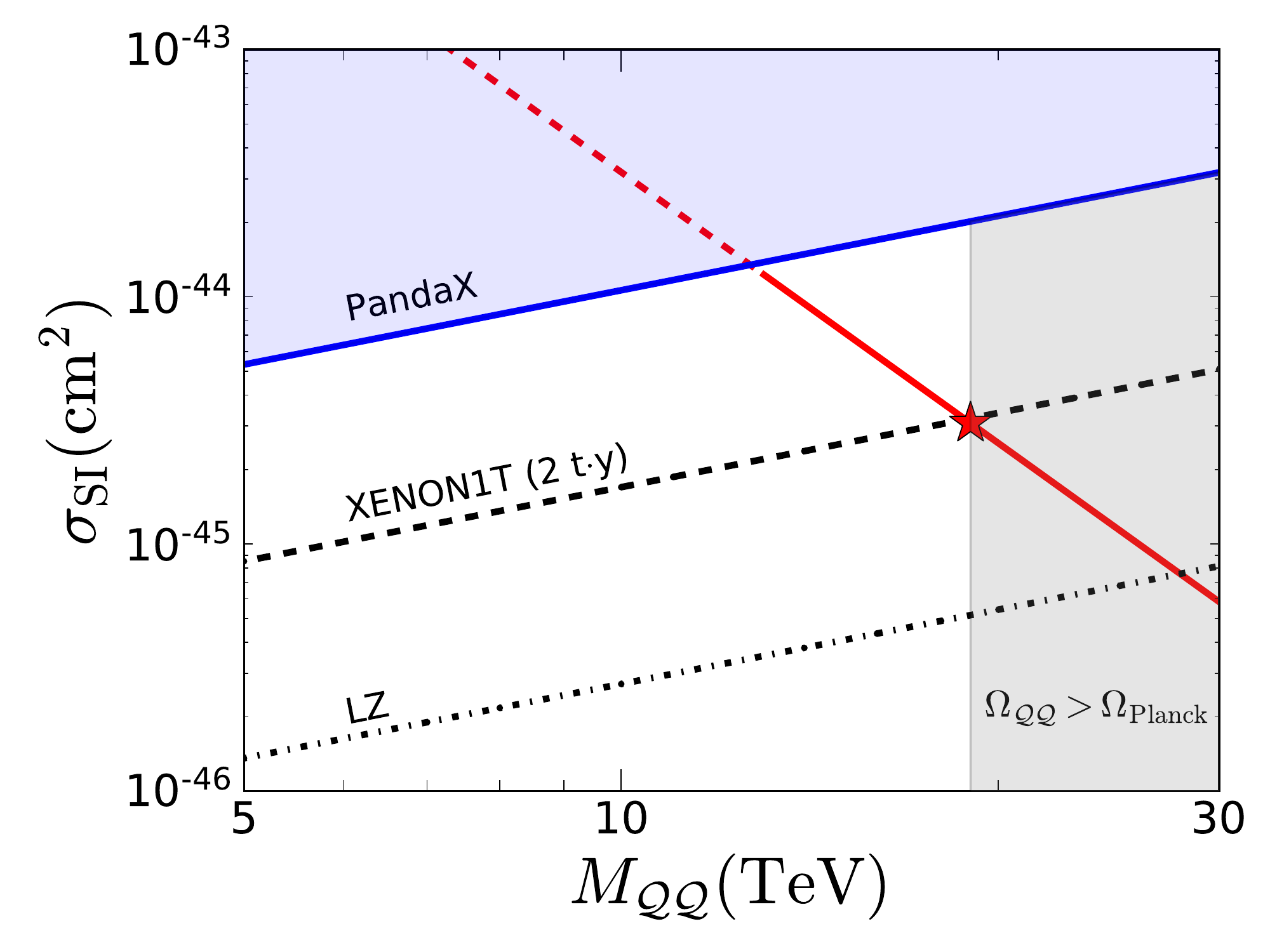}
\caption{The red line gives the spin independent cross section as a
  function of $M_{\mathcal{Q}\mathcal{Q}}=2M_\mathcal{Q}$. The star
  denotes the mass required for a thermal bound state DM
  (19~TeV). Smaller values can be probed by direct searches, the blue
  line gives the current bound, while the black (dashed, dotted and
  dot-dashed) lines represent future sensitivities.  }
\label{fig:spin-independent}
\end{figure}

Bound-state dark matter will impart nuclear recoil in underground dark
matter search experiments. The spin-independent direct detection
cross-section is given as
 \begin{equation}
   \sigma_{\text{SI}}\approx 2\times 10^{-45}\ \text{cm}^2\left(\frac{20\ \text{TeV}}{M_{\mathcal{Q}\mathcal{Q} }}\right)^6 \frac{\Omega_{\mathcal{Q}\mathcal{Q}}}{\Omega_{\text{Planck}}}\,,
\end{equation}
sharply correlated to the dark matter mass $M_{\mathcal{Q}\mathcal{Q}}=2M_\mathcal{Q}$, as
shown in the red line in Fig.~\ref{fig:spin-independent}.

{In this figure the star corresponds to the case where our bound-state
  DM makes up 100\% of the cosmological dark matter.
  In the presence of an additional dark matter particle, such as the
  axion, bound-state dark matter masses below 19 TeV can be envisaged,
  as indicated by the red line.
In this case their contribution to the relic density will be
correspondingly smaller, while the spin-independent cross section
would be correspondingly larger.}
The blue line represents the current limit of
PandaX~\cite{Cui:2017nnn}.
  The black (dashed, and dot-dashed) lines represent the future
  sensitivities expected at XENON1T~\cite{Aprile:2015uzo} and
  LZ~\cite{Akerib:2018lyp}. 
  On the other hand we note that, within the standard thermal
  cosmological scenario, DM masses above 19~TeV are ruled out by
  current observations by the Planck collaboration~\cite{Ade:2015xua}
  (gray band).

  Notice that the current LHC limit of 2~TeV (next section) implies
  that the cross section is always small enough so as to have the
  bound-state dark matter candidate reaching underground detectors.
  For more detailed discussion and general aspects of the cosmology of
  a stable colored relic see~\cite{Geller:2018biy}.

\section{Color octets at hadron colliders }
\label{sec:color-octets-at}

In our model the messengers of neutrino mass generation are the
colored constituents of bound-state dark matter. Given enough energy,
the \Q's are copiously pair produced at hadron colliders, through the
processes $q_i\bar{q}_i\to \mathcal{Q}\bar{\mathcal{Q}}$ and
$g g\to \mathcal{Q}\bar{\mathcal{Q}}$, and are expected to hadronize.
In contrast to WIMP dark matter scenarios, which engender only
missing-energy signals, the bound-state dark matter scenario gives
rise to very visible signals at hadron colliders, as they can form
either neutral or charged bound states~\cite{DeLuca:2018mzn}, e.g.
neutral hybrid states $\mathcal{Q}g$ (detected as neutral hadrons,
presumably stable) or charged $\mathcal{Q}q\bar{q}^\prime$ states, or
more exotic $\mathcal{Q}qqq$ states, expected to be long-lived on
collider time-scales.

Current LHC data place a limit to the fermion color octet
mass, $M_\mathcal{Q}> 2$ TeV \cite{CMS:2016ybj}.
Since the the cosmological relic abundance requires $M_{\mathcal{Q}}\approx 9.5$
TeV, this scenario offers an attractive benchmark for future collider
experiments beyond the energies attainable at the LHC. In fact, from
the estimate in Ref.~\cite{diCortona:2016fsn} one finds that a hadron
collider of at least 65~TeV center-of-mass energy would be required to
probe the full cosmologically allowed range of masses of our
bound-state DM model. This will allow a cross-check of the DM search
results of XENON1T, expected quite soon, in just one year or so.
Concerning the scalar messengers, we have two pairs of these, the
$\sigma_a\sim (\mathbf{8},1,0)$, which are singlets under
$\operatorname{SU}(2)_L$, and the $\eta_a$, which transform as weak
doublets, $\eta_a\sim (\mathbf{8},\mathbf{2},1/2)$.
As color octets, these would also be copiously produced at a hadron
collider of sufficient energy~\cite{Hayreter:2017wra}.  However, their
masses are expected to lie well above the reach of the LHC.
Moreover, in our model these scalars carry non-trivial $B-L$ charges,
see Table \ref{tab:bmlnur}.
This makes them relatively \textit{inert} with respect to the \sm
fermions. This, in addition to their heavy masses, makes them very
difficult to probe directly.

\section{Lepton Flavor Violation}
\label{sec:lept-flav-viol}

\begin{figure}[!h]
\includegraphics[scale=0.5]{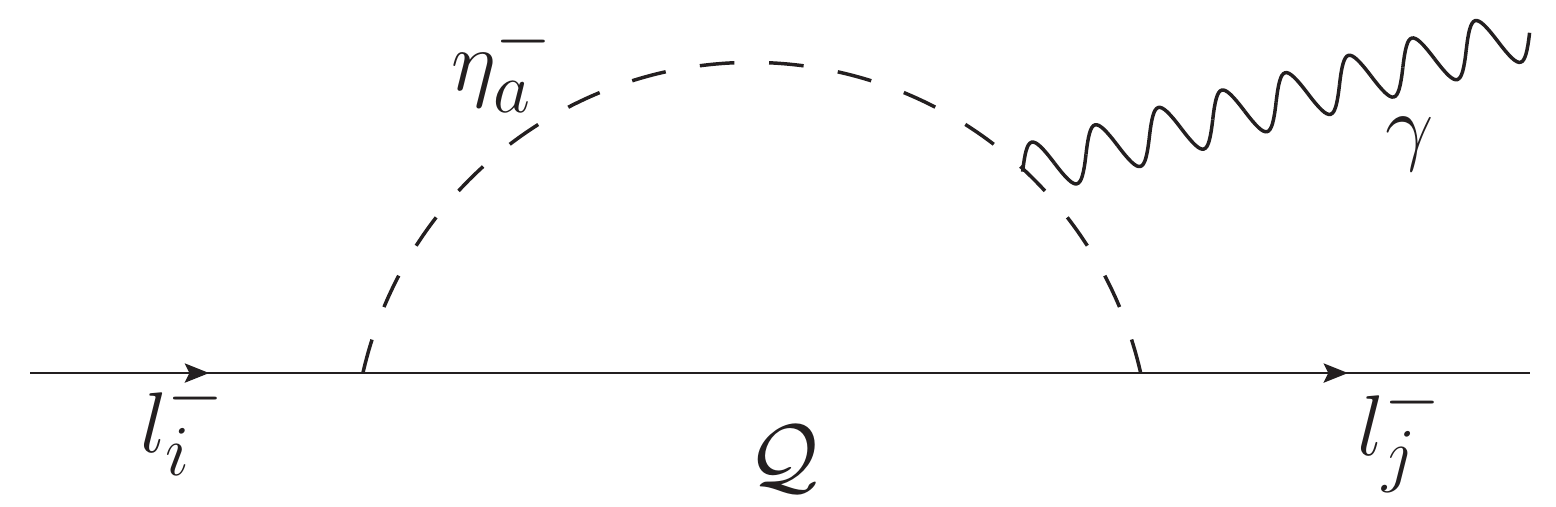}
\caption{Feynman diagram for the process $l_i\to l_j\gamma $}
\label{fig:muegamma}
\end{figure}

Our model may also lead to indirect virtual effects, such as charged
lepton flavor violation. For example, the Yukawa interactions in
Eq.~(\ref{eq:lag}) lead to radiative \lfv processes, as seen in
Fig.~\ref{fig:muegamma}, mediated by the charged scalar $\eta^+_a$.

The corresponding decay rate is given as~\cite{Lavoura:2003xp},
\begin{align}
\Gamma (l_i \to  l_j\gamma)=&\frac{e^2m_i^5}{16\pi}\Big[\sum_a N_c h^a_i h^{a*}_j \frac{i}{16\pi^2 M_{\eta_a^+}^2}\nonumber\\
& \left[\frac{-t^2\log t}{2(t-1)^4}+\frac{2t^2+5t-1}{12(t-1)^3}\right]\Big]^2\,,
\end{align}
with $t={M_{\mathcal{Q}}^2}/{M_{\eta^+}^2}$ and $N_c=8$. In the limit of heavy scalars, $t\to  0$, the decay width reads
\begin{align*}
\Gamma (l_i\to  l_j\gamma)=\frac{e^2m_i^5}{16\pi}\Big[\sum_a N_c h^a_i h^{a*}_j \frac{i}{16\pi^2 M_{\eta_a^+}^2}\left[-\frac{1}{12}\right]\Big]^2.
\end{align*}
One sees that the current experimental constraint
$\text{BR}(\mu\to  e\gamma)=\frac{\Gamma (\mu\to 
  e\gamma)}{\Gamma_\text{total}}< 5.7\times10^{-13}$, can be fullfilled provided
\begin{equation}
\left(\sum_a \frac{h^a_\mu h^{a*}_e}{M_{\eta_a^+}^2}\right)^2\leq 5.7\times 10^{-13}\frac{G_F^2}{\alpha_{\text{EM}}}\frac{768\pi}{N_c^2} ~,
\end{equation}
which leads to a relatively mild requirement,
\begin{equation}
\left|\sum_a h^a_\mu h^{a*}_e \left(\frac{50\,\mbox{TeV}}{M_{\eta_a^+}}\right)^2\right|\lesssim 1.5.
\end{equation}

\section{Summary and outlook}
\label{sec:summary-discussion}

We have proposed a consistent viable theory for the recently proposed
idea that the cosmological dark matter may be made up of stable
colored relics forming neutral hadronic bound states of QCD.
In our model we have taken up at face value the suggestion in
Ref.~\cite{DeLuca:2018mzn}, employing an exotic vector-like Dirac
color octet fermion \Q with mass below 10~TeV as the dark matter
constituent.
In our construction dark matter and neutrino mass generation both have
a common origin.
Our minimum particle content leads to two non-zero neutrino
masses, that can be associated to the solar and atmospheric scale.
Bound-state dark matter stability is directly associated with the
Dirac nature of neutrinos, and reflects the presence of an underlying
exact $B-L$ symmetry.
The scheme can account for both neutrino physics and dark matter
phenomena, within a consistent ultraviolet complete setup, free of
Landau poles up to the Planck scale, provided the scalars are heavy
enough.
Our model can be falsified relatively soon by dark matter searches,
and could also be cross-checked later by a next generation hadron
collider.
Variants of our construction may be envisaged, in which the dark
matter is bound by a new hypercolor
interaction~\cite{Mitridate:2017oky}, instead of QCD, as suggested in
Ref.~\cite{Reig:2017nrz}.

\begin{acknowledgments}

  We thank Martin Hirsch for very useful discussions.  Work supported
  by the Spanish grants FPA2017-85216-P and SEV-2014-0398 (MINECO),
  Sostenibilidad-UdeA, and by COLCIENCIAS through the Grants
  11156584269 and 111577657253.

\end{acknowledgments}

\bibliographystyle{utphys}
\bibliography{bibliography} 

\providecommand{\href}[2]{#2}\begingroup\raggedright\begin{thebibliography}{10}

\bibitem{DeLuca:2018mzn}
V.~De~Luca, A.~Mitridate, M.~Redi, J.~Smirnov, and A.~Strumia, ``{Colored Dark
  Matter},''
\href{http://arxiv.org/abs/1801.01135}{{\ttfamily arXiv:1801.01135 [hep-ph]}}.

\bibitem{Reig:2017nrz}
M.~Reig, J.~W.~F. Valle, C.~A. Vaquera-Araujo, and F.~Wilczek, ``{A Model of
  Comprehensive Unification},''
  \href{http://dx.doi.org/10.1016/j.physletb.2017.10.038}{{\em Phys. Lett.}
  {\bfseries B774} (2017) 667--670},
  \href{http://arxiv.org/abs/1706.03116}{{\ttfamily arXiv:1706.03116
  [hep-ph]}}.

\bibitem{Ma:2006km}
E.~Ma, ``{Verifiable radiative seesaw mechanism of neutrino mass and dark
  matter},'' {\em Phys.Rev.} {\bfseries D73} 077301,
  \href{http://arxiv.org/abs/hep-ph/0601225}{{\ttfamily arXiv:hep-ph/0601225
  [hep-ph]}}.

\bibitem{Hirsch:2013ola}
M.~Hirsch {\em et~al.}, ``{WIMP dark matter as radiative neutrino mass
  messenger},'' {\em JHEP} {\bfseries 1310} 149,
  \href{http://arxiv.org/abs/1307.8134}{{\ttfamily arXiv:1307.8134 [hep-ph]}}.

\bibitem{Merle:2016scw}
A.~Merle {\em et~al.}, ``{Consistency of WIMP Dark Matter as radiative neutrino
  mass messenger},'' \href{http://dx.doi.org/10.1007/JHEP07(2016)013}{{\em
  JHEP} {\bfseries 07} (2016) 013},
  \href{http://arxiv.org/abs/1603.05685}{{\ttfamily arXiv:1603.05685
  [hep-ph]}}.

\bibitem{Bonilla:2016diq}
C.~Bonilla, E.~Ma, E.~Peinado, and J.~W.~F. Valle, ``{Two-loop Dirac neutrino
  mass and WIMP dark matter},''
  \href{http://dx.doi.org/10.1016/j.physletb.2016.09.027}{{\em Phys. Lett.}
  {\bfseries B762} (2016) 214--218},
  \href{http://arxiv.org/abs/1607.03931}{{\ttfamily arXiv:1607.03931
  [hep-ph]}}.

\bibitem{Chulia:2016ngi}
S.~{Centelles Chuli{\'a}} {\em et~al.}, ``{Dirac Neutrinos and Dark Matter
  Stability from Lepton Quarticity},''
  \href{http://dx.doi.org/10.1016/j.physletb.2017.01.070}{{\em Phys. Lett.}
  {\bfseries B767} 209--213}, \href{http://arxiv.org/abs/1606.04543}{{\ttfamily
  arXiv:1606.04543 [hep-ph]}}.

\bibitem{deSalas:2017kay}
P.~F. de~Salas {\em et~al.}, ``{Status of neutrino oscillations 2017},''
  \href{http://arxiv.org/abs/1708.01186}{{\ttfamily arXiv:1708.01186
  [hep-ph]}}.

\bibitem{Aprile:2015uzo}
{\bfseries XENON} Collaboration, E.~Aprile {\em et~al.}, ``{Physics reach of
  the XENON1T dark matter experiment},''
  \href{http://dx.doi.org/10.1088/1475-7516/2016/04/027}{{\em JCAP} {\bfseries
  1604} no.~04, (2016) 027},
\href{http://arxiv.org/abs/1512.07501}{{\ttfamily arXiv:1512.07501
  [physics.ins-det]}}.

\bibitem{Schechter:1980gr}
J.~Schechter and J.~W.~F. Valle, ``{Neutrino Masses in SU(2) x U(1)
  Theories},'' \href{http://dx.doi.org/10.1103/PhysRevD.22.2227}{{\em Phys.
  Rev.} {\bfseries D22} (1980) 2227}.

\bibitem{Hirsch:2000ef}
M.~Hirsch {\em et~al.}, ``{Neutrino masses and mixings from supersymmetry with
  bilinear R parity violation: A Theory for solar and atmospheric neutrino
  oscillations},'' \href{http://dx.doi.org/10.1103/PhysRevD.62.113008,
  10.1103/PhysRevD.65.119901}{{\em Phys. Rev.} {\bfseries D62} (2000) 113008},
  \href{http://arxiv.org/abs/hep-ph/0004115}{{\ttfamily arXiv:hep-ph/0004115
  [hep-ph]}}.
[Erratum: Phys. Rev.D65,119901(2002)].

\bibitem{diaz:2003as}
M.~A. Diaz {\em et~al.}, ``{Solar neutrino masses and mixing from bilinear
  {R-parity} broken supersymmetry: Analytical versus numerical results},'' {\em
  Phys. Rev.} {\bfseries D68} 013009,
  \href{http://arxiv.org/abs/hep-ph/0302021}{{\ttfamily hep-ph/0302021}}.

\bibitem{hirsch:2004he}
M.~Hirsch and J.~W.~F. Valle, ``{Supersymmetric origin of neutrino mass},''
  {\em New J. Phys.} {\bfseries 6} 76,
  \href{http://arxiv.org/abs/hep-ph/0405015}{{\ttfamily hep-ph/0405015}}.

\bibitem{Farzan:2012sa}
Y.~Farzan and E.~Ma, ``{Dirac neutrino mass generation from dark matter},''
  \href{http://dx.doi.org/10.1103/PhysRevD.86.033007}{{\em Phys. Rev.}
  {\bfseries D86} (2012) 033007},
\href{http://arxiv.org/abs/1204.4890}{{\ttfamily arXiv:1204.4890 [hep-ph]}}.

\bibitem{Hehn:2012kz}
D.~Hehn and A.~Ibarra, ``{A radiative model with a naturally mild neutrino mass
  hierarchy},'' \href{http://dx.doi.org/10.1016/j.physletb.2012.11.034}{{\em
  Phys. Lett.} {\bfseries B718} (2013) 988--991},
\href{http://arxiv.org/abs/1208.3162}{{\ttfamily arXiv:1208.3162 [hep-ph]}}.

\bibitem{Manohar:2006ga}
A.~V. Manohar and M.~B. Wise, ``{Flavor changing neutral currents, an extended
  scalar sector, and the Higgs production rate at the CERN LHC},''
  \href{http://dx.doi.org/10.1103/PhysRevD.74.035009}{{\em Phys. Rev.}
  {\bfseries D74} (2006) 035009},
\href{http://arxiv.org/abs/hep-ph/0606172}{{\ttfamily arXiv:hep-ph/0606172
  [hep-ph]}}.

\bibitem{Bityukov:1997dh}
S.~I. Bityukov and N.~V. Krasnikov, ``{The Search for new physics by the
  measurement of the four jet cross-section at LHC and FNAL},''
  \href{http://dx.doi.org/10.1142/S0217732397002065}{{\em Mod. Phys. Lett.}
  {\bfseries A12} (1997) 2011--2028},
\href{http://arxiv.org/abs/hep-ph/9705338}{{\ttfamily arXiv:hep-ph/9705338
  [hep-ph]}}.

\bibitem{Bai:2018jsr}
Y.~Bai and B.~A. Dobrescu, ``{Collider Tests of the Renormalizable Coloron
  Model},''
\href{http://arxiv.org/abs/1802.03005}{{\ttfamily arXiv:1802.03005 [hep-ph]}}.

\bibitem{Cui:2017nnn}
{\bfseries PandaX-II} Collaboration, X.~Cui {\em et~al.}, ``{Dark Matter
  Results From 54-Ton-Day Exposure of PandaX-II Experiment},''
  \href{http://dx.doi.org/10.1103/PhysRevLett.119.181302}{{\em Phys. Rev.
  Lett.} {\bfseries 119} no.~18, (2017) 181302},
\href{http://arxiv.org/abs/1708.06917}{{\ttfamily arXiv:1708.06917
  [astro-ph.CO]}}.

\bibitem{Akerib:2018lyp}
{\bfseries LUX-ZEPLIN} Collaboration, D.~S. Akerib {\em et~al.}, ``{Projected
  WIMP sensitivity of the LUX-ZEPLIN (LZ) dark matter experiment},''
\href{http://arxiv.org/abs/1802.06039}{{\ttfamily arXiv:1802.06039
  [astro-ph.IM]}}.

\bibitem{Ade:2015xua}
{\bfseries Planck} Collaboration, P.~A.~R. Ade {\em et~al.}, ``{Planck 2015
  results. XIII. Cosmological parameters},''
  \href{http://dx.doi.org/10.1051/0004-6361/201525830}{{\em Astron. Astrophys.}
  {\bfseries 594} (2016) A13},
  \href{http://arxiv.org/abs/1502.01589}{{\ttfamily arXiv:1502.01589
  [astro-ph.CO]}}.

\bibitem{Geller:2018biy}
M.~Geller, S.~Iwamoto, G.~Lee, Y.~Shadmi, and O.~Telem, ``{Dark quarkonium
  formation in the early universe},''
\href{http://arxiv.org/abs/1802.07720}{{\ttfamily arXiv:1802.07720 [hep-ph]}}.

\bibitem{CMS:2016ybj}
{\bfseries CMS} Collaboration, C.~Collaboration,
``{Search for heavy stable charged particles with $12.9~\mathrm{fb}^{-1}$ of
  2016 data},''.

\bibitem{diCortona:2016fsn}
G.~Grilli~di Cortona, E.~Hardy, and A.~J. Powell, ``{Dirac vs Majorana gauginos
  at a 100 TeV collider},''
  \href{http://dx.doi.org/10.1007/JHEP08(2016)014}{{\em JHEP} {\bfseries 08}
  (2016) 014},
\href{http://arxiv.org/abs/1606.07090}{{\ttfamily arXiv:1606.07090 [hep-ph]}}.

\bibitem{Hayreter:2017wra}
A.~Hayreter and G.~Valencia, ``{LHC constraints on color octet scalars},''
  \href{http://dx.doi.org/10.1103/PhysRevD.96.035004}{{\em Phys. Rev.}
  {\bfseries D96} no.~3, (2017) 035004},
\href{http://arxiv.org/abs/1703.04164}{{\ttfamily arXiv:1703.04164 [hep-ph]}}.

\bibitem{Lavoura:2003xp}
L.~Lavoura, ``{General formulae for $f(1) \to f(2) \gamma$},''
  \href{http://dx.doi.org/10.1140/epjc/s2003-01212-7}{{\em Eur. Phys. J.}
  {\bfseries C29} (2003) 191--195},
\href{http://arxiv.org/abs/hep-ph/0302221}{{\ttfamily arXiv:hep-ph/0302221
  [hep-ph]}}.

\bibitem{Mitridate:2017oky}
A.~Mitridate, M.~Redi, J.~Smirnov, and A.~Strumia, ``{Dark Matter as a weakly
  coupled Dark Baryon},'' \href{http://arxiv.org/abs/1707.05380}{{\ttfamily
  arXiv:1707.05380 [hep-ph]}}.

\end{thebibliography}\endgroup
\end{document}